\documentclass[12pt]{article}

\usepackage{graphicx}
\usepackage{amssymb}
\usepackage{amsmath}

\numberwithin{equation}{section}

\begin{document}
\baselineskip=16pt
\begin{titlepage}
\begin{flushright}
{\small IC/2008/011}
\\[-1mm]{\small RIKEN-TH-126}
\end{flushright}
\begin{center}
\vspace*{12mm}

{\Large\bf%
Quark Mass Deformation of Holographic Massless QCD%
}\vspace*{10mm}

Koji Hashimoto$^1$\footnote{e-mail:
{\tt koji@riken.jp}},
Takayuki Hirayama$^{2,3}$\footnote{e-mail:
{\tt hirayama@phys.cts.nthu.edu.tw}},
Feng-Li Lin$^3$\footnote{e-mail:
{\tt linfengli@phy.ntnu.edu.tw}} and
Ho-Ung Yee$^4$\footnote{e-mail:
{\tt hyee@ictp.it}}
\vspace*{4mm}

${}^1$
{\it Theoretical Physics Laboratory, RIKEN, Saitama 351-0198, Japan}\\[1mm]
${}^2$
{\it Physics Division, National Center for Theoretical Sciences, Hsinchu 300, Taiwan}\\[1mm]
${}^3$
{\it Department of Physics, National Taiwan Normal University, Taipei 116, Taiwan}\\[1mm]
${}^4$ {\it The Abdus Salam International Center for Theoretical
Physics, Strada Costiera 11, 34014, Trieste, Italy}
\end{center}
\vspace*{10mm}

\begin{abstract}\noindent%
We propose several quark mass deformations of the holographic model
of massless QCD using the D4/D8/$\overline{\rm D8}$-brane
configuration proposed by Sakai and Sugimoto. The deformations are
based on introducing additional D4- or D6-branes away from the QCD D4
branes. The idea is similar to  extended
technicolor theories, where the chiral symmetry breaking by
additional D-branes is mediated to QCD to induce non-zero quark
masses. In the D-brane picture as well as the holographic dual
gravity description, the quark and the pion masses are generated by
novel worldsheet instantons with finite area. We also derive the
Gell-Mann-Oakes-Renner relation, and find the value of the chiral
condensate in the Sakai-Sugimoto model.

\end{abstract}

\end{titlepage}

\newpage


\section{Introduction}

AdS/CFT correspondence~\cite{Maldacena:1997re} has provided us with
a new avenue to compute observables in strongly interacting gauge
theories via weakly coupled gravitational descriptions. Even though
the correspondence is still a conjecture, many nontrivial
computations have been performed and checked to be consistent. One
of the most important applications, and also challenges, would be to
analyze real QCD in Nature using the AdS/CFT correspondence.

Recently, there proposed a holographic model of QCD by Sakai and
Sugimoto, where the non-Abelian chiral symmetry and its spontaneous
breaking is geometrically realized via a D4/D8/$\overline{\rm D8}$
configuration~\cite{Sakai:2004cn}. The model predictions for the
meson spectrum and the couplings between them compare well with
experiments with a good accuracy. Baryons in this model are also
investigated in~\cite{Hong:2007kx}. Moreover, this holographic model
has an ability of predicting possible glue-ball interactions to
mesons as studied in~\cite{Hashimoto:2007ze}. There are many other
QCD observables that have been analyzed in this model,
which show that the model is a good approximation to real QCD at
least in low energy, large $N_c$-limit.

Despite its success in many aspects, this model has an apparent
shortcoming that the pions are massless. This is because the quarks
are massless from the construction. Therefore one of the important
questions is to find a deformation of the theory which corresponds
to introducing bare quark masses. Once we identify the deformation,
we should be able to show that the Gell-Mann-Oakes-Renner (GOR)
relation \cite{GellMann:1968rz} is satisfied, and moreover, we can
compute the value of the chiral condensate of the Sakai-Sugimoto model
through this relation.
Pion dynamics would be the most important in holographic QCD because
the Sakai-Sugimoto model approaches real QCD in the low energy limit,
thus the identification of the quark mass deformation in the model is
indispensable.

The chiral quarks appear as the lowest massless modes of the open
string stretching between intersecting $N_c$ D4-branes and $N_f$
D8-branes (or $\overline{\rm D8}$-branes) in the model. Therefore
one would naively expect that the quarks become massive if one can
realize the situation where the D8 and $\overline{\rm D8}$-branes do
not intersect with the D4-branes. Some of the
authors~\cite{Hashimoto:2007fa}
studied such local deformations near the D4-branes (which do not
deform the asymptotic configuration away from the D4-branes) to realize
the pion mass, but the quarks in this set-up are suggested to be still
massless. This is in fact consistent with the proposed
relation~\cite{Gubser:1998bc} between the classical value of action
in the bulk and the partition function in CFT for the
AdS$_5$/CFT$_4$ correspondence, since the value of bulk field at the
AdS boundary is related with the strength of coupling, such as the
mass, in CFT. There are attempts to
include D8-$\overline{\rm D8}$-brane tachyon in the effective action to
explicitly break the chiral symmetry \cite{Casero:2007ae}. However,
tachyon action is not reliable as there is no consistent truncation
of the tachyon theory to low energy. Furthermore, possible relation
to the original Sakai-Sugimoto brane configuration is not obvious.

In this work, we propose
deformations of the Sakai-Sugimoto model
which correspond to the introduction of the quark mass.
Our deformations have additional D-branes away
from the original confining $N_c$ D4-branes, which still have a
reasonable field theory
interpretation. The original Sakai-Sugimoto model is restored once these
additional D-branes are moved to spatial infinity in the bulk.
In the field theory view-point, our idea is rather
similar to the technicolor model~\cite{mini}, so our construction
can be thought of as a holographic realization of technicolor model.
In the minimal technicolor
model, we have an additional sector to the original QCD, which contains
new techni-quarks interacting via new technicolor interactions.
The quarks/techni-quarks are massless at UV, but the chiral symmetry (the
electroweak symmetry in this case) is spontaneously broken by a techni-quark condensate driven by the strong
technicolor gauge interactions. To realize QCD quark mass from this breaking,
we need to introduce extended technicolor interactions between quarks and techni-quarks, in addition to the usual weak interactions~\cite{ETC}.
This setup is actually close to the holographic dual of extended type
technicolor model by applying the D-brane configuration of
Sakai-Sugimoto model in \cite{Hirayama:2007hz}.
This extended gauge group can often be realized in GUT type construction,
where quarks and techni-quarks are in a same multiplet, and the GUT gauge symmetry is broken to QCD and technicolor at some high scale. Through
off-diagonal massive gauge bosons, QCD quarks get explicit bare mass term
from the techni-quark condensate. {}From the view-point of QCD, we have
a realization of explicit chiral symmetry breaking mass for the quarks.

To realize this idea in our set-up, we
introduce additional $N'$ D4-branes (we call them D4$'$-branes)
which are parallel to, but separated from the original $N_c$ D4-branes of QCD gauge symmetry. The GUT gauge symmetry is  $SU(N_c+N')$ when
D4$'$-branes are on top of the original D4-branes, and it is broken to $SU(N_c)\times SU(N')$ by a Higgs mechanism via separating D4$'$ from D4.
Massive off-diagonal gauge bosons appear as strings between D4 and D4$'$.
We add $N_f$ D8/$\overline{\rm  D8}$-branes as in \cite{Sakai:2004cn}
to introduce massless quarks and techni-quarks,
which are charged under the chiral symmetry $U(N_f)\times U(N_f)$.
By assuming a strong technicolor $SU(N')$ dynamics on the D4$'$-branes,
which replaces D4$'$-branes with the Witten geometry,  the chiral
symmetry is spontaneously broken by adjoining D8/$\overline{\rm
D8}$-branes there, as in the Sakai-Sugimoto
model but with technicolor instead of QCD.  This breaking will be
mediated to the QCD sector of $N_c$ D4-branes by, for example, the
massive gauge bosons coming from open strings between the D4 and
D4$'$-branes. This is expected to induce the bare
masses for the QCD quarks.

{}From the Feynman graphs that would induce quark masses from a
techni-quark condensate, we can easily
identify the corresponding string worldsheets that mediate this phenomenon in
our D-brane configuration in  flat space. We consider a closed loop
surrounded by D4-D8-D4$'$-$\overline{\rm D8}$-branes (see Fig.\ref{d2})
and a disk {\it worldsheet instanton} whose boundary is
ending on this closed loop. It is responsible for the extended technicolor interactions coupling quarks to techni-quarks.
As we go to the strongly coupled field theory, or weakly coupled gravity regime,
we keep track of how this worldsheet looks like in the gravity description,
as we will study in Sec.~\ref{ws}. In the gravity dual description,
where both D4 and D4$'$-branes are replaced by the near horizon
geometry of multi center solution,
we have a closed loop with the probe D8-branes only, since D8 and
$\overline{\rm D8}$-branes are smoothly connected at two throats
created by the D4 and the
D4$'$-branes (see Fig.\ref{d4}). The previous worldsheet
instanton in the weak coupling picture is
now ending on this closed loop, and is understood as a leading order
contribution, since this is a planar diagram with one
boundary in large $N$ gauge theory~\cite{'t Hooft:1973jz}.
We will show that this worldsheet instanton amplitude indeed
induces the lowest mass perturbation that one may expect in the low energy
chiral Lagrangian of pions. We will also show that
the GOR relation is satisfied.
This will be presented in detail in Sec.~\ref{ws}.

Since the chiral symmetry is broken spontaneously in view of the whole
GUT theory, there still exist massless Goldstone bosons.
These modes will be localized around D4$'$ throat when the techni-quark condensate scale is much bigger than the QCD scale.
In the gravity picture, they would correspond to the Wilson line on D8
from the asymptotic boundary to the D4$'$ throat and again to the
boundary. 
We have to take a limit to decouple these
Nambu-Goldstone bosons by scaling $N'\rightarrow\infty$ first and
putting D4$'$-branes at far UV region from the D4-branes, while keeping
finite the mass of pions associated with the chiral symmetry
breaking realized by $N_c$ D4-branes. Also, it seems complicated  to
analyze multi-center solutions of D4-branes explicitly.
It is difficult to realize flavor dependent quark mass in this set-up, too.

We then consider introducing $N'$ D6-branes instead of the
D4$'$-branes, which is free of these difficulties. This deformation keeps the essential idea of introducing $N'$ D4-branes, but has great advantages of computability without taking any subtle limits. Therefore, we can
compute the quark mass as well as the pion mass more rigorously,
and consequently we can estimate the value of the chiral
condensate of the Sakai-Sugimoto model via the GOR relation.
We will also see that the quark masses derived from our results with the pion
mass as an input will be around $6$ MeV, which happens to be quite
close to the real QCD. The numerical value of the computed chiral
condensate is close to the results of lattice QCD.
This will be presented in Sec.~\ref{sqm}.

In the next section, we first give the field theoretical interpretation
of introducing the $N'$ D4-branes and describe the Feynman graphs which
induce the quark masses. Then we study the
corresponding string worldsheet instantons  both in the D-brane configuration in  flat space and in the holographic dual gravity description, and
estimate the quark and pion masses. After explaining difficulties of precise
computations, we instead introduce $N'$ D6-branes and study the corresponding
interpretation in QCD side in Sec.~\ref{sqm}, where we carry out explicit
computations of quark and pion masses.
We show in detail that the up and down quarks have masses around $6$
MeV in our model. It is important to stress that we can
also realize  flavor dependent quark masses with our D6-branes. Since our idea
has a field theoretical interpretation which can be applied to a wide range of holographic QCD models, we apply our idea to a holographic
approach of QCD proposed by
Kruczenski {\it et.al}~\cite{Kruczenski:2003uq} in Sec.\ref{d4d6case}.
We give a detail about how our idea works in their model.
In Sec.~\ref{cd}, we conclude with discussions.


\section{Quark mass deformation and worldsheet instantons}
\label{ws}

In the Sakai-Sugimoto
model with $N_c$ D4, $N_f$ D8 and $N_f$ $\overline{\rm D8}$-branes,
QCD gauge bosons are realized on the D4-branes and the chiral flavor
symmetry $U(N_f)_L\times U(N_f)_R$ is realized as the gauge symmetry on
the $N_f$ D8 and $N_f$ $\overline{\rm D8}$-branes. The left (right)
handed massless quarks are localized at the intersection between the
D4-branes and the D8- ($\overline{\rm D8}$-) branes.
In the holographic dual description,
the near horizon geometry of the D4-branes develops a throat
which naturally explains confinement.  The chiral symmetry
breaking in QCD is nicely realized
since the D8 and $\overline{\rm D8}$-branes are connected
smoothly at the throat.

If we introduce additional D4-branes (D4$'$-branes) which are parallel to
but separated from the original D4-branes, the chiral symmetry is also
broken by the D4$'$-branes in the holographic dual description of strong
technicolor dynamics on the D4$'$-branes. Since there
are massive gauge bosons from the open strings stretching between the D4
and D4$'$-branes, the effect of the
chiral symmetry breaking at the D4$'$-branes
will be mediated into QCD sector on the D4-branes and the quark mass
terms are resultantly induced. We will pursue this idea in this section.


\subsection{Quark mass deformation}

\begin{figure}[t]
\begin{center}
\includegraphics[width=8.2cm]{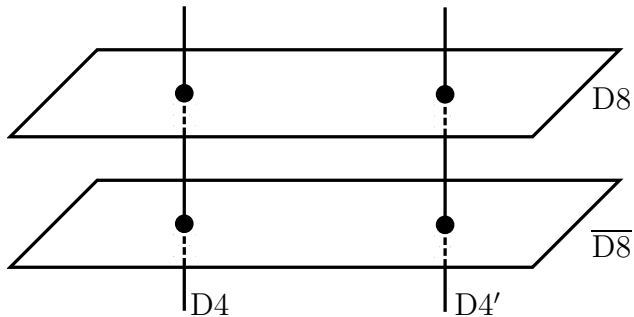}
\put(-164,0){D4}
\put(-64,0){D4$'$}
\put(-12,78){D8}
\put(-12,22){$\overline{\rm D8}$}
\caption{The D-brane configuration}
\label{d1}
\end{center}
\end{figure}

We first describe the D-brane configuration and study the low energy
theory on the D-branes. The D-brane configuration which we now consider
consists of D4, D8 and $\overline{\rm D8}$-branes (see Fig.\ref{d1})
and is summarized in the table:
\begin{center}
\begin{tabular}{c|c|c|c|c|c|c|c|c|c|c}
& 0&1 &2  &3 &4  &5 &6  &7 &8  &9
\\
\hline
$N_c$ D4 & $\circ$ &$\circ$ &$\circ$  &$\circ$ &$\circ$  & &  & &  &
\\
\hline
$N_f$ D8/$\overline{\rm D8}$ & $\circ$ &$\circ$ &$\circ$  &$\circ$ & &$\circ$ &$\circ$  &$\circ$ &$\circ$  &$\circ$
\\
\hline
$N'$ D4$'$ & $\circ$ &$\circ$ &$\circ$  &$\circ$ &$\circ$  & &  & &  &
\end{tabular}
\end{center}
where $N_c$ D4-branes are localized at $x^5=\cdots=x^9=0$ and $N'$
D4-branes are localized at $x^5\neq 0$ and $x^6=\cdots=x^9=0$. The D8 and
$\overline{\rm D8}$-branes are localized at $x^4=0$ and
$x^4=\delta \tau/2$ where $x^4$ direction is compactified by a
supersymmetry breaking $S^1$ with the periodicity
$x^4\sim x^4+\delta\tau$. The compactification scale determines the scale
of this system and $M_{KK}\equiv 2\pi/\delta\tau \sim 1$ GeV in the
Sakai-Sugimoto model~\cite{Sakai:2004cn}.

\begin{figure}[t]
\begin{center}
\begin{minipage}{7.5cm}
\begin{center}
\includegraphics[height=4.5cm]{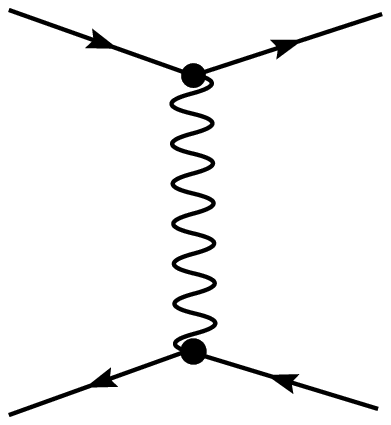}
\put(-112,110){$q_L$}
\put(-112,10){$q_R$}
\put(-10,110){$Q_L$}
\put(-10,10){$Q_R$}
\put(-86,60){$W_\mu$}
\caption{The four fermi interaction mediated by massive gauge boson $W_\mu$.}
\label{w1}
\end{center}
\end{minipage}
\hspace{4ex}
\begin{minipage}{7.5cm}
\begin{center}\hspace*{-4ex}
\includegraphics[width=4.5cm]{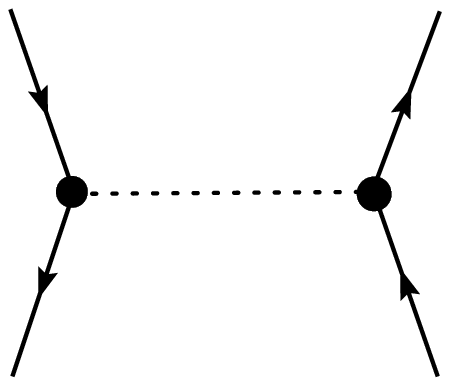}
\put(-116,100){$q_L$}
\put(-116,4){$q_R$}
\put(-22,100){$Q_L$}
\put(-22,4){$Q_R$}
\put(-70,64){$T$}
\caption{The four fermi interaction mediated by massive scalar field $T$.}
\label{t1}
\end{center}
\end{minipage}
\end{center}
\end{figure}

The $N'$ D4-branes realize another massless QCD with $SU(N')$ gauge
symmetry as a low energy effective theory in addition to the
massless QCD with $SU(N_c)$ in the Sakai-Sugimoto model. Thus the
massless modes are $SU(N_c)\times SU(N')$ gauge bosons and the two
sets of left handed and right handed chiral massless quarks $(q_L,
q_R)$ and $(Q_L,Q_R)$ which are fundamentally charged under
$SU(N_c)$ and $SU(N')$ respectively. The massless modes are
summarized in the table:
\begin{center}
$\begin{array}{c|cccc}
  & SU(N_c) & SU(N') & U(N_f)_L & U(N_f)_R\\
  \hline
  A_\mu & {\rm adj.}
  \\
  q_L & \Box & & \Box
  \\
  q_R & \Box & & & \Box
  \\
  \hline
  A'_\mu & & {\rm adj.}
  \\
  Q_L & & \Box &  \Box
  \\
  Q_R & & \Box & & \Box
\end{array}$
\end{center}
Since $Q_L$ ($Q_R$) come from the open strings stretching between
$N'$ D4 and $N_f$ D8 ($\overline{\rm D8}$)-branes, the flavor
symmetry is still $U(N_f)_L\times U(N_f)_R$ and there are massive
modes which connect $(q_L,q_R)$ and  $(Q_L,Q_R)$. They are the modes
originated from the open strings stretching between the D4 and
D4$'$-branes or D8 and $\overline{\rm D8}$-branes. The lightest modes
from these two strings are massive gauge bosons $W_\mu$ and complex
scalar fields $T$ whose charges are summarized in the table:
\begin{center}
$\begin{array}{c|cccc}
  & SU(N_c) & SU(N') & U(N_f)_L & U(N_f)_R\\
 \hline
  W_\mu & \Box & \Box
  \\
  T & & & \Box &\Box
\end{array}$
\end{center}
Their masses are computed from the distance between the D4- and
D4$'$-branes or the D8- and the $\overline{\rm D8}$-branes.\footnote{
The field $T$ has been referred to as a ``tachyon'' since it is from a
string connecting the D8 and the $\overline{\rm D8}$-branes. However,
when those branes are separated as in the present case, it is massive.
In the low energy limit $l_s\rightarrow 0$ on the D4-branes, the mass of
$T$ diverges.
}
The quarks have a
gauge and Yukawa interaction with $W_\mu$ and $T$ fields and then
there are two Feynman graphs in Fig.\ref{w1}-\ref{t1}
which generate the four fermi interactions
$\bar{q}_Lq_R\bar{Q}_LQ_R+$h.c.~ mediated by the massive gauge or the
scalar fields.

\begin{figure}[t]
\begin{center}
\begin{minipage}{7.5cm}
\begin{center}
\includegraphics[height=4.5cm]{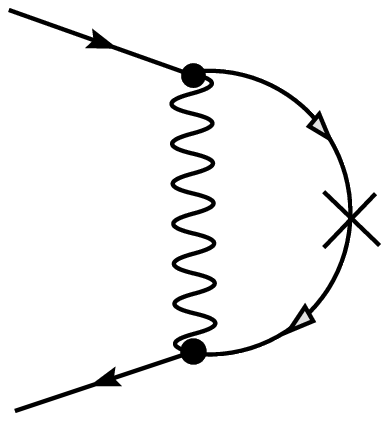}
\put(-112,110){$q_L$}
\put(-112,10){$q_R$}
\put(0,58){$\langle\bar{Q}_LQ_R\rangle$}
\put(-86,60){$W_\mu$}
\end{center}
\end{minipage}
\hspace{4ex}
\begin{minipage}{7.5cm}
\begin{center}\hspace*{-4ex}
\includegraphics[width=4.5cm]{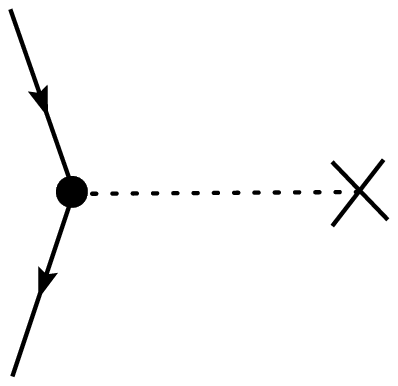}
\put(-116,116){$q_L$}
\put(-116,4){$q_R$}
\put(0,60){$\langle\bar{Q}_LQ_R\rangle$}
\put(-70,68){$T$}
\end{center}
\end{minipage}
\caption{The quark mass is generated from the condensate
$\langle\bar{Q}_LQ_R\rangle$.}
\label{t2}
\end{center}
\end{figure}

At the low energy both $SU(N_c)$ and $SU(N')$ gauge symmetries become
confined and the chiral symmetry is broken by the condensate
$\langle\bar{q}_Lq_R\rangle$ and $\langle\bar{Q}_LQ_R\rangle$. Then
by picking up the condensate $\langle\bar{Q}_LQ_R\rangle$, the quark
$(q_L,q_R)$ becomes massive $m_q\propto \langle\bar{Q}_LQ_R\rangle$.
See Fig.~\ref{t2}.
This mechanism is quite similar to the mechanism of inducing the
quark masses in extended technicolor theories.

{}From the low energy effective theory on D-branes in the weak
coupling picture, we can estimate the quark masses $m_q$. The
effective theory on the D4-branes is essentially a five dimensional
theory and quarks are localized at spatially separated 4 dimensional
surfaces where they intersect with the D8-branes. Thus the
calculation involves Kaluza-Klein (KK) reduction along the $S^1$
with a few steps and a similar calculation has been done in a
context of brane world scenario in~\cite{Mirabelli:1997aj} (see
appendix \ref{app0} for the calculations). We just mention that the
amplitude in Fig.\ref{w1} is suppressed by $\exp[-\pi M_W/M_{KK}]$
since the W-boson with the mass $M_W$ must propagate from $\tau=0$
to $\tau=\delta\tau/2=\pi/ M_{\rm KK}$. We can similarly compute the
Feynman graph Fig.\ref{t1} and obtain the same suppression
$\exp[-\pi M_W/M_{KK}]$: the D8/$\overline{\rm D8}$-brane tachyon
$T$ with mass $\frac{1}{2\pi\alpha'}{\delta\tau\over 2}$ propagating
horizontally from the D4$'$ to the D4 by distance $U_0=2\pi\alpha'
M_W$. In string theory, the string worldsheet which represents these
two graphs is identical. It is $st$-channel duality on the string
worldsheet which manifests in these two Feynman graphs. Then it is
in fact easier to estimate the masses by studying the corresponding
string worldsheets.


\subsection{Worldsheet instanton}

\begin{figure}[t]
\begin{center}
\begin{minipage}{7.5cm}
\begin{center}
\includegraphics[width=7cm]{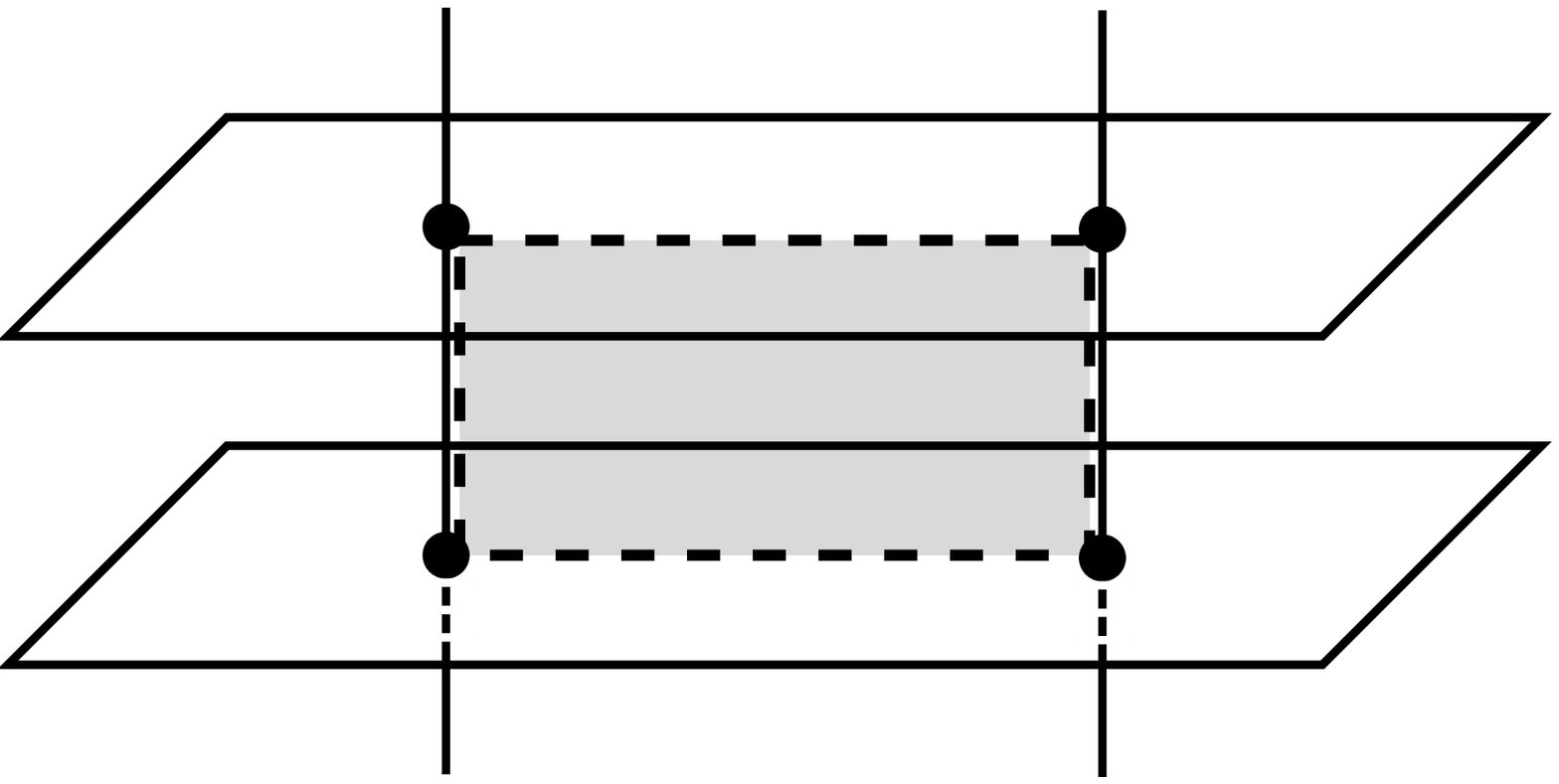}
\caption{The shaded region denotes a worldsheet instanton. A dotted line
 is a closed loop surrounded by the D4-D8-D4$'$-$\overline{\rm D8}$ branes.}
\label{d2}
\end{center}
\end{minipage}
\hspace{4ex}
\begin{minipage}{7.5cm}
\begin{center}\hspace*{-4ex}
\includegraphics[width=7cm]{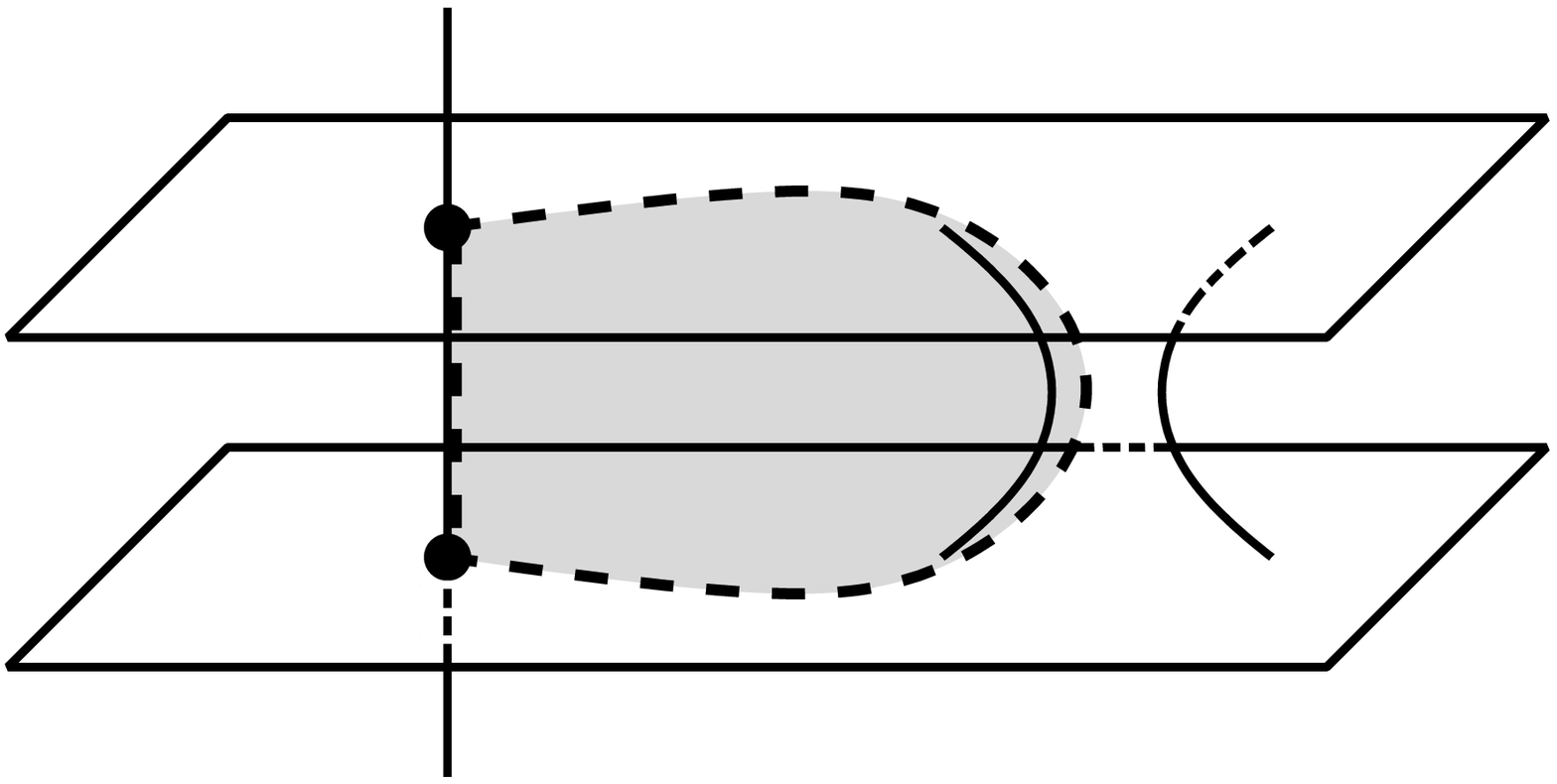}
\caption{The shaded region denotes a worldsheet instanton. The throat is
 developed at the location of D4$'$-branes in the geometry.}
\label{d3}
\end{center}
\end{minipage}
\end{center}
\end{figure}

The string worldsheets which correspond to the Feynman graphs in
Fig.\ref{w1}-\ref{t1} can be easily identified. The chiral quarks are
localized on the intersections of D-branes of different types and open strings stretching between D-branes of the same type induce massive gauge or scalar fields. Also the graphs
in Fig.\ref{w1}-\ref{t1} are tree graphs and we conclude that the
corresponding string worldsheets are disk worldsheets whose boundary
are on the closed loop surrounded by D4-D8-D4$'$-$\overline{\rm D8}$
branes
and which have the four boundary twist vertex operators at the
intersections of the D-branes. Since these worldsheets are localized in
time direction, these are worldsheet instantons which are
denoted by a shaded region in Fig.\ref{d2}.
\footnote{When we have $N_f>1$ number of D8- and $\overline{\rm
D8}$-branes, we have other worldsheets ending on, for example,
D4$^{(i)}$-D8-D4$'{}^{(j)}$-D8 branes. These worldsheets do not give
rise to the quark masses (and the pion masses).}

The quark masses are induced by the chiral condensate
$\langle\bar{Q}_LQ_R\rangle$ realized by strongly interacting $SU(N')$
gauge dynamics. This dynamics can be captured from a weakly coupled
holographic dual description by replacing the
D4$'$-branes by the near horizon
geometry and still keeping the D4-branes as probes in
order to keep the $SU(N_c)$ in a weak coupling regime.
The geometry is given by Witten~\cite{Witten:1998zw} and the throat is
developed at the location
of the D4$'$-branes to connect the D8- and $\overline{\rm D8}$-branes
smoothly which describes the spontaneous chiral symmetry breaking.

In this geometry, the worldsheet instantons which we like to evaluate are
the ones whose boundary is on the D4- and the D8-branes in Fig.\ref{d3}
with two twist vertex operators which describe the quarks. By evaluating
the amplitude in a saddle point approximation, we can show that
the quarks become massive since the amplitude is given as
\begin{eqnarray}
 S_{\rm instanton} &= c \int d^4x ~{\rm tr}~
\bar{q}_Lq_R \exp[-S_{NG}] +{\rm h.c.} .
\end{eqnarray}
Here the trace is taken over the flavor $U(N_f)$ indices and $S_{NG}$ is
the classical value of the Nambu-Goto action.
The Hermitian conjugate comes from the oppositely
oriented worldsheets, and $c$ is a constant factor. We have to
evaluate $S_{NG}$ in the curved geometry of
\cite{Witten:1998zw} which is
\begin{eqnarray}\label{metric-1}
&& ds^2
= \bigg( {U \over R'} \bigg)^{3/2} \big( dx_4^2 + f(U) d\tau^2 \big)
 + \bigg( {R' \over U} \bigg)^{3/2}
 \bigg( {d U^2 \over f(U)} + U^2 d \Omega_4^2 \bigg),
 \label{aho}
\\
&& e^\phi = g_s \bigg( {U \over R'} \bigg)^{3/4}, \quad
 F_4=\frac{2\pi N'}{V_4}\epsilon_4,
 \quad f(U) = 1 - {{U'}_{\rm KK}^{3} \over U^3} ,
 \\
&& {R'}^3 = \pi g_s N' l_s^3, \hspace{3ex}
 M_{KK} = \frac{3{U'}_{KK}^{\frac{1}{2}}}{2{R'}^{\frac{3}{2}}} ,
\end{eqnarray}
where $U \geq U'_{KK}$ is the radial direction transverse to the
D4$'$-branes,
$g_s$ is the string coupling, $l_s=\sqrt{\alpha'}$ is the string length,
$V_4$ and $\epsilon_4$ are the volume and line element of $S_4$.

Let the probe $N_c$ D4-brane be placed at $U=U_0$. Since the minimal size
worldsheet instanton extends along $U$ and $\tau$ directions, we can
evaluate
\begin{eqnarray}
 S_{NG} = \frac{1}{2\pi\alpha'} \int_0^{\delta\tau/2} d\tau
 \int_{U'_{KK}}^{U_0} dU
 \sqrt{g_{\tau\tau}g_{UU}}
 = \frac{U_0-U'_{KK}}{2\alpha' M_{KK}}
 = \frac{\pi M_W}{M_{KK}} \label{SNG-1}
\end{eqnarray}
in the approximation $U_0 \gg U'_{\rm KK}$, where $M_W$ is the mass
of the massive gauge boson $W_\mu$ computed as $ M_W =
{U_0}/(2\pi\alpha')$.
We finally obtain
\begin{eqnarray}
  m_q =c \exp\left[-\frac{\pi M_W}{M_{KK}}\right] .
\end{eqnarray}
The exponential suppression factor is the same as that in the field theory
calculation which supports our identification of the
worldsheet instantons
with the four-fermi Feynman graphs. The constant factor $c$ depends on
the string length, the string coupling
and the chiral condensate $\langle\bar{Q}_LQ_R\rangle$, and then is an
important coefficient. However it is difficult to compute $c$ in the
curved geometry and this leads us to the study of
introducing D6-branes instead
of the D4$'$-branes in the next section.


\subsection{Pion mass}

We found that the quark becomes massive due to the chiral symmetry
breaking caused by the $N'$ D4-branes. Thus we should be able to
see that the pions
become massive and the mass should satisfy the GOR relation.

In the low energy limit, both $SU(N_c)$ and $SU(N')$ become strongly
coupled and the holographic dual description gives a weakly coupled
description in terms of gauge invariant operators, {\it i.e.}
mesons. In doing
that, we need a multi-center solution of the
D4-branes and its near horizon
geometry. However the explicit metric is not known and in the next
section we study a more computable model. Thus in this subsection we only
give a qualitative discussion on how the pions are shown to be massive in
the holographic dual description.

\begin{figure}[t]
\begin{center}
\includegraphics[width=8.2cm]{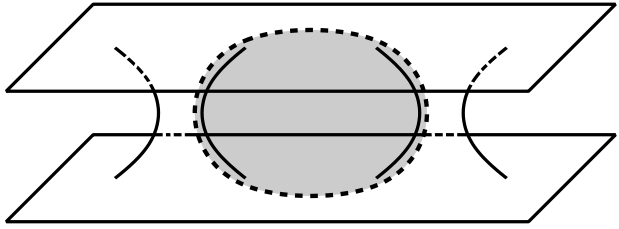}
\caption{The shaded region denotes a worldsheet instanton. The two
 throats are developed at the location of D4 and D4$'$-branes.}
\label{d4}
\end{center}
\end{figure}

The geometry of the two-center solution of D4-branes has two throats at the
location of the D4- and the D4$'$-branes respectively. See Fig.\ref{d4}.
The geometry is roughly obtained by gluing two metrics \eqref{aho}.
Thus we again have the worldsheet instantons denoted by the shaded
region in the same figure. The amplitude of this worldsheet
instantons is given by
\begin{eqnarray}
  S_{\rm instanton} \propto \int d^4x \frac{1}{g_s} {\rm Ptr} \exp
  \left[ -S_{NG} + i\oint dz A_z \right] +{\rm h.c.} ,
\end{eqnarray}
where we have included a boundary coupling to $A_z$ and $z$
parameterizes the boundary of the worldsheet instanton. The factor $1/g_s$ is
introduced because the worldsheet is a disk. Since the pion wave function is
localized at the D4 throat~\cite{Sakai:2004cn}, we have
\begin{eqnarray}
  {\rm Ptr} \exp \left[ i\oint dz A_z \right] = {\rm tr}~ U ,
\quad U \equiv \exp \left[i2\pi(x)/f_\pi\right],
\end{eqnarray}
after substituting into $A_z$ the pion wave function. Then we obtain
\begin{eqnarray}
  S_{\rm instanton} = m_\pi^2 f_\pi^2 \;{\rm tr} (U+U^\dagger) ,
\quad
  m_\pi^2 \propto \frac{1}{g_s} \exp\left[ -S_{NG} \right] \propto m_q
\end{eqnarray}
because the exponent $\exp [ -S_{NG}]$ is roughly same as that of the
quark mass.
We can see the GOR relation is naturally satisfied.
This is
the lowest mass deformations in the pion chiral Lagrangian which should
appear by the quark mass perturbations to QCD.

\begin{figure}[t]
\begin{center}
\begin{minipage}{7.5cm}
\begin{center}
\includegraphics[width=5cm]{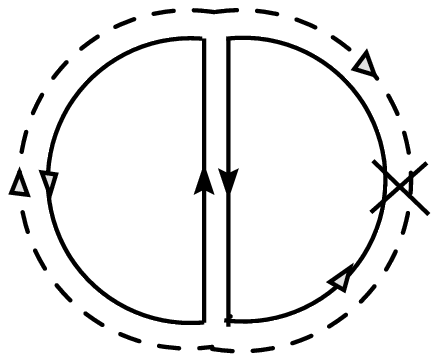}
\end{center}
\end{minipage}
\hspace{4ex}
\begin{minipage}{7.5cm}
\begin{center}\hspace*{-4ex}
\includegraphics[width=5cm]{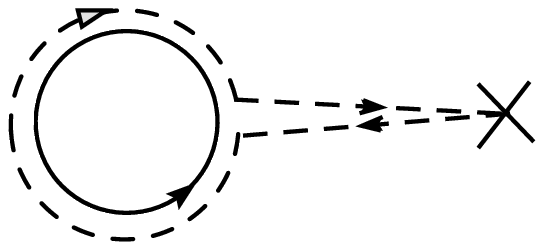}
\end{center}
\end{minipage}
\caption{The Feynman diagrams responsible for quark masses can be written
 in the double-line notation, which shows that the flavor line (dashed
 lines) form a closed single loop. Solid lines are for the color indices.}
\label{t3}
\end{center}
\end{figure}

Before closing this section, we give some interesting observations. The
holographic description is reliable in the large $N_c$ (and $N'$) and
large 't Hooft coupling, and a genus zero string worldsheet would
correspond
to a planar diagram in a large $N$ QCD proposed by 't Hooft. The Feynman
graphs using the double-line notation in a large $N$ QCD is shown in
Fig.~\ref{t3}.
It is clear from this that these Feynman diagrams are included in a
planar diagram with one boundary. On the other hand, the worldsheet
instanton is a disk amplitude which is nothing but a planar diagram
with one boundary. This gives a consistency that we identify the four-fermi Feynman graphs with the worldsheet instantons.

The worldsheet instanton does not contribute to the
Dirac-Born-Infeld (DBI) action $S_{\rm DBI}$ for the
D8-branes, {\it i.e.}
the total action  $S_{\rm D8}$ is
\begin{eqnarray}
 S_{\rm D8} &= S_{\rm DBI} + S_{\rm instanton} .
\end{eqnarray}
There must not be a mass term for the pions from $S_{\rm DBI}$ at the
leading order in large $N$. We can easily see this from
the fact that the pion comes from a KK zero mode in the gauge
field on D8-branes along the extra dimension. Since the gauge fields
always appear through field strengths in the DBI action, the pion
$\pi(x)$ always appears with four-dimensional derivatives, i.e.
$\partial_\mu \pi(x)$, and cannot have a mass term.

In this section, we focused on the quark $(q_L, q_R)$ and the pion in
the $SU(N_c)$ sector. Because the D4$'$-branes realize another $SU(N')$
gauge theory, we have $(Q_L, Q_R)$ and another pion in $SU(N')$ sector.
Since the chiral symmetry is still spontaneously broken even though we
introduced the D4$'$-branes, one combination of the two pions are still
massless Nambu-Goldstone bosons. In order to decouple these modes
from the QCD sector, we need to take a limit which realizes the pion
decay constant for this unwanted pion infinite by taking $N'$
to the infinity as well as taking the distance from D4-branes infinite
while keeping the pion mass we computed finite. Since we can not
explicitly compute $S_{\rm instanton}$, the actual realization of
this limit is not clear.

Therefore as we will explain in the next section, we consider another
deformation by instead introducing D6-branes which keeps the idea in this
section while needs no limit, {\it i.e.} the chiral symmetry is
explicitly broken by the D6-branes.


\section{Quark mass deformation by D6-branes and Pion mass}
\label{sqm}

\setcounter{footnote}{0}

In this section, we propose another way to introduce quark masses
to the Sakai-Sugimoto model. In the approach of this section,
the evaluation of the quark and pion mass terms is
more tractable. First we explain the idea of introducing a probe
D6-brane ending on the D8- and the $\overline{\rm D8}$-branes, instead
of the D4$'$-branes. Then, we compute the worldsheet instanton explicitly
to obtain the quark masses, pion mass and the chiral condensate.
Finally, we present a numerical evaluation of those values,
for a tentative comparison with experimental or lattice QCD's results.

\subsection{Configuration of the D6 ending on D8 and
$\overline{\rm\bf D8}$}

As seen in the previous section we can introduce the quark mass term
if we have a 1-cycle in the brane configuration, because the worldsheet
instantons are possible. {}From this point of view, we do not necessarily
use the D4$'$-branes and instead introduce $N'$
D6-branes which are
separated away from the D4-branes and end on the D8-branes and the
$\overline{\rm D8}$-branes
(each D6-brane can end on different D8- and $\overline{\rm D8}$-branes)
. The orientation of the additional probe D6-brane is
shown in the table.\footnote{
To have a stable configuration of the
D6-brane, the D6-brane is curved when the D4-brane is replaced by
its geometry.
In the appendix \ref{appA}, we find a consistent curved shape
of the D6-brane in the background, by solving the equations of
motion of the D6-brane effective action.
}
\begin{center}
\begin{tabular}{c|c|c|c|c|c|c|c|c|c|c}
& 0&1 &2  &3 &4  &5 &6  &7 &8  &9
\\
\hline
$N_c$ D4 & $\circ$ &$\circ$ &$\circ$  &$\circ$ &$\circ$  & &  & &  &
\\
\hline
$N_f$ D8
$\overline{\rm D8}$
& $\circ$ &$\circ$ &$\circ$  &$\circ$ & &$\circ$ &$\circ$
 &$\circ$ &$\circ$  &$\circ$
\\
\hline
$N_f$ D6 & $\circ$ &$\circ$ &$\circ$  &$\circ$ &$\circ$  & &$\circ$  &
$\circ$ &  &
\end{tabular}
\end{center}

This brane-ending-on-brane configuration is represented by a stable
spike solution on the D8-branes. This means
that the D8-branes and the $\overline{\rm D8}$-branes
are smoothly connected with each other by the smooth spike
(volcano-shaped)~\cite{Callan:1997kz}, and
thus there is no twist operator insertion at the end points of the D6.
We can obtain a worldsheet instanton whose shape is almost the same as
that of the case introducing the D4$'$-branes.\footnote{
This instanton is similar to the one considered in \cite{BL}
where QCD instanton is studied. We thank J.~Sonnenschein for pointing
this out.
}

Let us summarize our field theoretical understanding on the D-branes
with emphasis on the differences from the D4$'$-brane case.
\begin{itemize}
\item
The D6-brane
can be treated as
a probe {\bf ($N'\ll N_c$)},
hence it does not modify the background geometry.
The D6-brane is a classical solution on the D8-branes (or the
$\overline{\rm D8}$-branes), and it is a part of the probe flavor
D-branes. Thus, on
the contrary to the previous section,
the background is still given by the $N_c$ D4-branes.

\item
The introduction of the D6-brane breaks the chiral symmetry, since the D6-branes end on the D8 and $\overline{\rm D8}$-branes and thus connect the D8 and $\overline{\rm D8}$-branes. The breaking is explicit, not spontaneous, since the D6-branes extend along $(x^6,x^7)$ infinitely and the fields localized on the D6-branes,  which would be Nambu-Goldstone bosons, do not have a finite kinetic term in the four dimensions.

\item
{}From the theory on the D8-branes, the D6-branes are interpreted as
monopoles with the charge $(1,-1)$ under the chiral symmetry for a
single flavor case $U(1)_L\times U(1)_R$. Thus the source for the Dirac
monopole explicitly breaks the axial symmetry.
For D6-branes not on top of each other, all the axial symmetry is
broken, which corresponds to the situation where quark masses are not
coincident with each other.

\item
The field theoretical understanding of how the effects of chiral
     symmetry breaking on the D6-brane sector are mediated into the QCD
     sector on the D4-branes is essentially the same. Here we have
     massive scalar fields (instead of massive gauge bosons) from the
     open strings between the D4-branes and the D6-branes which, for
     example, play the role of the mediators.

\item
We have an easy way to realize flavor-dependent quark masses by
shifting the location of each D6-brane independently. Another way
(which can also apply for the D4$'$-branes)
is to shift the D8-brane from the anti-podal points, but the D8-brane
configuration away from the anti-podal point is only numerically
known~\cite{Aharony:2006da} which makes the estimate more difficult.

\end{itemize}

\subsection{Quark mass from worldsheet instanton}

Let us first compute the quark mass term. At this stage
we keep the D4-branes to be probes, that is,
the spacetime is still flat. (In the next subsection we replace them
with a curved
geometry.) The mass generation mechanism by the worldsheet instantons
is quite the same as in the previous section, see Fig.~\ref{d3}.
The throat on the right hand side of the figure
is now
$N'$ spiky D6-branes ending on the D8-branes
and the $\overline{\rm D8}$-branes.
At the end points of the D6-brane, there is no insertion of a vertex
operator
because the D6 and the D8 join smoothly there.
Resultantly, the worldsheet instanton amplitude should be given by
\begin{eqnarray}
\frac{{\cal N}N'}{g_s l_s}e^{-S_{\rm NG}}
\int \!
d^4x \; \bar{q}_{\rm L}  q_{\rm R} .
\end{eqnarray}

The classical worldsheet action $S_{\rm NG}$ is just the area of the
worldsheet times string tension in flat space, so we have $S_{\rm
NG} = (1/(2\pi\alpha')) U_0 (\pi/M_{\rm KK})=\pi M_W/M_{\rm KK}$
where $U_0$ is the length of an open string stretching between the
D4- and D6-branes and then $M_W$ is the mass of modes from the open
string. The factor $N'$ comes since the boundary of worldsheet
instantons can be on $N'$ different D6-branes. The front factor
$1/g_s$ is for a disk amplitude, and the $1/l_s$ factor should be
provided from the normalization of the quark vertex operator, to
make a sensible dimensionality of the amplitude. The dimensionless
number ${\cal N}$ is a normalization factor from the quark vertex
operators, the quark kinetic terms and the fluctuations around our semi-classical worldsheet.

For the multi-flavor case, by shifting the position of each D6-brane
respectively,  we can introduce quark masses depending on flavors.
For each flavor labeled by $i$, we can associate the location
$U=U_0^{(i)}$ of coincident $N'$ D6-branes (then the total number of
the D6-branes is $N'N_f$). Then obviously, the quark mass term is a
sum of each instanton, $\sum_i \int\! d^4x\; \bar{q}^{(i)}_{\rm L}
(m_q)_{ij}  q^{(j)}_{\rm R}$, with the following diagonal quark mass
matrix
\begin{eqnarray}
(m_q)_{ij}
=
{\cal N}N'
 \delta_{ij}\frac{2\pi M_{\rm KK}}{g_{\rm YM}^2}
\exp\left[
-\frac{\pi M^{(i)}_W}{M_{\rm KK}}
\right] ,
\label{quarkmasst}
\end{eqnarray}
where $M_W^{(i)}=U_0^{(i)}/(2\pi\alpha')$.

\subsection{Pion mass, GOR relation and chiral condensate}

Next,
we study the low energy limit in the holographic dual by replacing
the D4-branes by their geometry (Witten's geometry) and
compute the pion mass term. Since there is a nontrivial 1-cycle on the
D8-brane, the worldsheet instanton gives us a pion mass term (see
Fig.~\ref{d4}). The mechanism is completely analogous to the previous
section. In the following, we present explicit evaluation of the
worldsheet instanton amplitude. We will adopt some
crude approximations for obtaining the results
because a curved background makes the explicit evaluation quite difficult.

The worldsheet instanton amplitude is given as
\begin{eqnarray}
S_{\rm instanton} =
N'
\frac{1}{g'_s}
\frac{1}{(2\pi)^3l_s^4}
\int \!\sqrt{-\det g}\; d^4x \; e^{-S_{\rm NG}}
\left[
{\rm Ptr}\exp\left[-i\oint A_z dz \right]
+ {\rm c.c.}
\right] ,
\label{inst6}
\end{eqnarray}
where the trace is taken over the flavor indices.
To obtain this expression, we made the
following approximations as we will explain step by step.

The classical worldsheet action $S_{\rm NG}$ is evaluated in the same
way computed in the previous section, {\it i.e.},
see (\ref{SNG-1}),  so  $S_{\rm NG} = \pi M_W/M_{\rm KK}$.

Let us explain the front factor $1/(g'_s (2\pi)^3l_s^4)$ in (\ref{inst6}).
First, we notice that the worldsheet instanton is
wrapping the minimal cycle whose zero mode is 4-dimensional.
The worldsheet instanton cannot move along the direction of the
$S^4$: the boundary should be on the D6-brane whose closest point to the
D4-branes is just a point on the $S^4$ (see appendix \ref{appA}
for details).
So the instanton amplitude should be proportional to a D3-brane tension
${\cal T}_{\rm D3}$, not ${\cal T}_{\rm D8}$.\footnote{
Along the $S^4$
directions transverse to the worldsheet instanton configuration, the
string fluctuation follows Dirichlet boundary condition rather than
Neumann. This effectively reduces the zero mode integral of the
vertex insertion at the boundary of the worldsheet instanton. The
number of the transverse directions is 4 in the worldvolume of the
D8-branes. Taking into account the $z$ direction along which the
worldsheet boundary is elongated, we need to
introduce the factor ${\cal T}_{\rm D3}$ in front of the worldsheet
instanton amplitude, rather than just $1/g_s$.}
The D3-brane tension is given by the front
factor $1/(g'_s (2\pi)^3l_s^4)$.

Note that the string coupling constant should be the effective
coupling constant $g'_s$
obtained by including the effect of the background dilaton,
$1/g'_s = (1/g_s) \left(R/U_{\rm KK}\right)^{3/4}$ (see the classical
solution (\ref{aho})
but remove the prime in the corresponding quantities.).
In getting this expression we made a crude (but reasonable)
approximation that the $U$-dependence of the dilaton can be
approximated by the value at
$U\sim U_{\rm KK}$. This is because
the pion wavefunction is localized around
$U_{\rm KK}$ so that most of the contribution for the integral over $U$ would come from the region $U\sim U_{\rm KK}$,
even though the worldsheet itself is elongated in the region
$U_{\rm KK} < U < U_0$. Furthermore,
the invariant volume $\sqrt{\det g}$ in the
$(x^0,x^1,x^2,x^3)$ directions was inserted in (\ref{inst6})
for consistency. Using the metric expression (\ref{metric-1}),
the four-volume is given as
$\sqrt{-\det g} = \left({U_{\rm KK}}/{R}\right)^3$.

We can calculate the instanton amplitude (\ref{inst6}) using
the relations between the gravity and the gauge theory quantities, namely,
\begin{eqnarray}
R^3 = \frac{g_{\rm YM}^2 N_c l_s^2}{2 M_{\rm KK}}, \quad
 U_{\rm KK} = \frac29 g_{\rm YM}^2 N_c M_{\rm KK} l_s^2,\quad
g_s = \frac{g_{\rm YM}^2}{2 \pi M_{\rm KK}l_s},
\label{grel}
\end{eqnarray}
then we arrive at
\begin{eqnarray}
 S_{\rm instanton} =
 \frac{2 N'}{3^{9/2}\pi^2}
g_{\rm YM} N_c^{3/2}M_{\rm KK}^4 \exp\left[ -\frac{\pi M_W}{M_{\rm
KK}} \right] \int d^4x\; ({\rm tr} U + {\rm tr} U^\dagger).
\end{eqnarray}
This is the pion mass term generated by the worldsheet instanton in the
curved background. Again, we found an expression well-known in chiral
perturbation theory.
Substituting the quark mass formula (\ref{quarkmasst}),
this can be re-written as
\begin{eqnarray}
 S_{\rm instanton} =
\frac{1}{3^{9/2}\pi^3}
g_{\rm YM}^3 N_c^{3/2}M_{\rm KK}^3
{\cal N}^{-1}m_q
\int d^4x\; ({\rm tr} U + {\rm tr} U^\dagger).
\label{uudagger}
\end{eqnarray}

Expanding $U = \exp [2i\pi(x)/f_\pi]$ to the second order in the pion
field $\pi(x)$, we obtain the GOR relation
\begin{eqnarray}
 m_\pi^2 =
\frac{4g_{\rm YM}^3 N_c^{3/2}M_{\rm KK}^3}{3^{9/2}\pi^3{\cal N}}
\frac{1}{f_\pi^2} m_q.
\label{gor}
\end{eqnarray}
The chiral condensate computed in our model is, therefore,
\begin{eqnarray}
 \langle \bar{q}q\rangle =
\frac{2}{3^{9/2}\pi^3}{\cal N}^{-1}g_{\rm YM}^3 N_c^{3/2}M_{\rm KK}^3 .
\label{chiralcond}
\end{eqnarray}
This is the expression for the chiral condensate
in the Sakai-Sugimoto model.
\footnote{
An interesting feature of this
chiral condensate is that, if ${\cal N}=1$,
it is independent of $N_c$ if it is written
with 'tHooft coupling $\lambda= g_{\rm YM}^2 N_c$. It sounds consistent
with Coleman-Witten argument \cite{Coleman:1980mx}.
The $N_c$ dependence of the chiral condensate found here is
a little different from what was found in \cite{Kruczenski:2003uq}.}

Note that, as we stated, we made crude approximations in evaluating
the coordinate dependence of the background dilaton and metric.
But it is nontrivial that all the $l_s$ dependence disappears at the
end, which may signal a consistency of the approximation.

\subsection{Flavor-dependent quark masses and numerical results}

\setcounter{footnote}{0}

In this subsection we consider the flavor dependent quark masses by
placing the D6-branes at different points.  The  worldsheet
instanton ends in part on the D6-brane, and let us consider the
transverse scalar field $\Phi$ on the D6-branes. The separation of
the D6-branes can be encoded in the worldsheet boundary interaction
as a condensation of the transverse scalar field $\Phi$. Because the
D6-branes are made of the spike of the D8- and $\overline{\rm
D8}$-branes, this $\Phi$ can be at the same time regarded as a
transverse scalar field on the D8-$\overline{\rm D8}$-branes, too.
So, in total, together with the usual boundary interaction for the
gauge fields on the D8-branes, the boundary interaction is
\begin{eqnarray}
{\rm Ptr} \exp\left[-i\oint A_z dz - \int_b \Phi dz\right]
\label{bsca}
\end{eqnarray}
The integration region $b$
is a period (in the worldsheet boundary) where the worldsheet ends on
the D6-branes.\footnote{
We don't have the factor $i$ in front of the scalar field
coupling, because the scalar field is associated with the instanton
which is Euclideanized, and we will see the consistency below.}
 We can parametrize the scalar field as
$2\pi\alpha'\Phi = {\rm diag} (U_0^{(1)}, U_0^{(2)}, \cdots)$ where
$U_0^{(i)}$ denotes the location of the $N'$ D6-branes\footnote{This
is the location of the tip of the smeared D6-brane cone, see
appendix \ref{appA} for the shape of the D6-brane. The tip is given
by $\theta_1=0$.} for the $i$-th flavor. If we neglect the dynamical
fluctuation of the scalar field, then this is just a constant
matrix.

The vertex insertion of the gauge field $A_z dz$ on the boundary of the
worldsheet instanton is located almost
around the tip of the cigar (due to the localized distribution
of the pion wave function found in \cite{Sakai:2004cn}),
that is far away from the region where the
transverse scalar $\Phi$  is inserted (where the D6-brane is present).
Therefore, the path-ordering in (\ref{bsca})
is effectively approximated by
the following ordering
\begin{eqnarray}
 {\rm tr}\left[{\rm P}\exp \left[-\Phi \int_{b} dz\right]
\cdot {\rm P} \exp\left[-i\oint A_z dz \right]
\right].
\label{phia}
\end{eqnarray}

Note that the integral $\int_b$ is performed only along the worldsheet
boundary on the D6-branes. This means
\begin{eqnarray}
\exp\left[
 -\Phi \int_{b}dz \right]
=
\exp\left[
 -\Phi \frac{\pi}{M_{\rm KK}}\right]
=
\exp\left[
-{\rm diag}
\left(
\frac{\pi M_W^{(1)}}{M_{\rm KK}},
\frac{\pi M_W^{(2)}}{M_{\rm KK}},\cdots
\right)
\right]
=\frac{g_{\rm YM}^2}{2\pi M_{\rm KK}}m_q.
\label{qmf}
\end{eqnarray}
Note that we realize a non-abelian generalization of the worldsheet area
$S_{\rm NG} \simeq \pi M_W/M_{\rm KK}$ 
in this manner. $S_{NG}=0$ in this notation.
If instead one uses the
minimal value of the $S_{\rm NG}$ which is $\pi M_W^{(1)}/M_{\rm
KK}$, then one should parameterize the transverse scalar field as
its deviation from $U^{(1)}$, as $2\pi\alpha'\Phi = {\rm diag} (0,
U_0^{(2)}-U^{(1)}, U_0^{(3)}-U_0^{(1)},\cdots)$ to get the correct
non-Abelian expression for the worldsheet boundary interaction.
Using this expression,  we obtain the following term induced by the
worldsheet instanton:
\begin{eqnarray}
 S_{\rm instanton} = \frac{1}{3^{9/2}\pi^3}
g_{\rm YM}^3 N_c^{3/2}M_{\rm KK}^3{\cal N}^{-1}
\int d^4x\; {\rm tr} \left[
m_q (U +  U^\dagger)
\right].
\label{ws1}
\end{eqnarray}
This form, ${\rm tr}[m_q(U+U^\dagger)]$,
is in fact what is usually expected in chiral perturbation theory.

In the two-flavor case, we have $m_q = {\rm diag} (m_u, m_d)$, so
\begin{eqnarray}
 {\rm tr}
\left[
m_q (U+U^\dagger)
\right]
=
2 (m_u+m_d)
\left(1 - \frac{1}{f_\pi^2}  \pi_a(x)^2 + {\cal O}(\pi(x)^4)\right).
\end{eqnarray}
where the component definition is $\pi(x) = \pi_a(x)\sigma_a/2$.
So at this leading order estimate
the mass of $\pi^0$ is equal to the mass of $\pi^\pm$, as in the usual
chiral perturbation theory.
Using the expression \cite{Sakai:2004cn} for the pion decay constant
$f_\pi=g_{\rm YM} N_c M_{\rm KK}/(3\sqrt{6}\pi^2)$, (\ref{ws1}) gives a
pion mass term
\begin{eqnarray}
  m_\pi^2 = 2(m_u + m_d)
\frac{2\pi g_{\rm YM} M_{\rm KK}}{3\sqrt{3} N_c^{1/2} {\cal N}}.
\label{pionmass}
\end{eqnarray}

Just for an illustration, we present a numerical evaluation of the
relation (\ref{gor}).\footnote{ Readers are advised that the values
presented below are just for illustration, and should not be taken
seriously, as we made a crude approximation in evaluating the
worldsheet instanton in the curved background, and we are working in
the leading order in large $N$ and large 'tHooft coupling
expansion.} Sakai and Sugimoto \cite{Sakai:2004cn} deduced numerical
values as $g_{\rm YM} = 2.35, M_{\rm KK} = 949 \; {\rm [MeV]}$. The
first came from the equation $\kappa = \lambda N_c/(216 \pi^3) =
0.00745$ with $N_c=3$. Then, with $m_\pi = 140$ [MeV] as an input,
the mass formula (\ref{pionmass}), with ${\cal N}=1$ as an
assumption, gives
\begin{eqnarray}
 m_u+m_d = 6.29 \; {\rm [MeV]}.
\end{eqnarray}
Experimental results shown in the particle data book are
$m_u = 1.5-4.0 \; {\rm [MeV]}$, $m_d = 4-8 \; {\rm [MeV]}$.
Surprisingly,
our result is very close to the observed
value of the up/down quark mass.

The difference between $\pi_0$ and $\pi_\pm$ can be seen in higher-order
chiral Lagrangian in chiral perturbation theory. In appendix \ref{appB},
we give
a computation of two worldsheet instantons, to get the higher-order
terms.

The chiral condensate (\ref{chiralcond}) is numerically evaluated as
\begin{eqnarray}
 \langle \bar{q}q\rangle =
(299 \; {\rm [MeV]})^3,
\end{eqnarray}
which is in quite good agreement with values obtained in
quenched/unquenched lattice simulation,
$\langle \bar{q}q\rangle \simeq (2.5\times 10^2 \; {\rm [MeV]})^3$.


\section{Application to the holographic QCD of flavor D6-branes}
\label{d4d6case}

Our idea is based on a field theoretical picture inspired by extended
technicolor theories. The technicolor sector is coupled with the QCD
sector via massive gauge-bosons of the extended gauge group, and the
condensation of the techni-quarks $Q$ gives rise to a quark mass
term through the induced four-Fermi coupling. Therefore, this
mechanism can be applied to other holographic models of QCD. In this
section we demonstrate this by
applying our idea to the D4-D6 model of Kruczenski {\it et.al}
\cite{Kruczenski:2003uq}. The model consists of $N_c$ D4-branes
giving rise to the
curved geometry and $N_f$ flavor D6-branes which are introduced as
probes. In fact the model can describe massive quarks, as the
D6-branes can be shifted away from the D4-branes. We will see that
our technicolor D4$'$-branes can shift the D6-branes further and the
quarks get extra masses because of this.


The D-brane configuration of the model in~\cite{Kruczenski:2003uq}
consists of the D4 and D6-branes as in the table:
\begin{center}
\begin{tabular}{c|c|c|c|c|c|c|c|c|c|c}
& 0&1 &2  &3 &4  &5 &6  &7 &8  &9
\\
\hline $N_c$ D4 & $\circ$ &$\circ$ &$\circ$  &$\circ$ &$\circ$  & &
& &  &
\\
\hline $N_f$ D6 & $\circ$ &$\circ$ &$\circ$  &$\circ$ &  &$\circ$
&$\circ$  &$\circ$ &  &
\end{tabular}
\end{center}
 The radial directions in $(x^5,x^6,x^7)$ and in $(x^8,x^9)$ from
the D4-branes are denoted as $\lambda$ and $r$ respectively, and the
D6-brane configuration in the D4-brane geometry is solved in their
paper. The asymptotic behavior for $\lambda \gg \lambda_c$,
($\lambda_c$ is a some large value), is
\begin{eqnarray}
 r \sim r_{\infty} + \frac{c}{\lambda} ,
\end{eqnarray}
where $r_\infty$ is the asymptotic distance between the D4- and the
D6-branes
which is related to the bare quark mass $m_q$ as
\begin{eqnarray}
 m_q = \frac{r_\infty}{2\pi l_s^2},
\end{eqnarray}
and $c$ is related to the value of condensate\footnote{ $U_{KK}=1$
is used in~\cite{Kruczenski:2003uq}.}
\begin{eqnarray}
 \langle\bar{q}q\rangle \simeq
\frac{g_{YM}^2 N_c^2 M_{KK}^3}{U_{KK}^2} c
 \hspace{3ex}
 \left(U_{KK} = \frac{2}{9}g_{YM}^2 N_c M_{KK}l_s^2\right)
\end{eqnarray}
The mass of the pion, which is a pseudo Nambu-Goldstone boson associated
with the breaking of the rotation symmetry of the $(x^8,x^9)$-plane,
is also computed in their paper
\begin{eqnarray}
 m_\pi^2 \sim \frac{m_q M_{KK}}{g_{YM}^2N_c} .
\end{eqnarray}

We now introduce our technicolor $N'$ D4$'$-branes which are parallel
to, but separated from the $N_c$ D4-branes along $\lambda$ with
$r=0$ (so that $U=\lambda$). Let's say that the D4-branes are sitting at
$\lambda=0$ and the D4$'$-branes are at $\lambda=\lambda_0$. We first
treat the D4-branes as probes and consider the effects from confining
D4$'$ dynamics. Assuming the distance $U_0=\lambda_0$ between the D4
and D4$'$-branes is much larger than $\lambda_c$, i.e. $U_0\gg
\lambda_c$, the value of $r$ at the position of the D4-brane
$\lambda=0$ {\it induced by the presence of D4$'$-brane} would be
\begin{eqnarray}
 r \sim r_\infty + \frac{c'}{U_0}\quad,
 \end{eqnarray}
where $c'$ denotes the contribution from the D4$'$-branes, {\it i.e.}
\begin{eqnarray}
 \langle\bar{Q}Q\rangle \simeq
\frac{g_{YM}^2 N'^2 M_{KK}^3 }{{U'}_{KK}^2}c'
 \hspace{3ex}
 \left(U'_{KK} = \frac{2}{9}g_{YM}^2 N' M_{KK}l_s^2\right)
\end{eqnarray}
For the probe D4-branes before their gravity background is considered,
this
would serve as its new asymptotic distance $r_\infty$ from the $D6$
branes. It is shifted from the original $r_\infty$ by
$\frac{c'}{U_0}$. We then take the gravity background for our D4-branes,
and we have
\begin{eqnarray}
 r \sim r_\infty + \frac{c'}{U_0}+ \frac{c}{\lambda} ,
 \label{asy2}
\end{eqnarray}
for $\lambda_c \ll \lambda \ll U_0$.

Therefore we see that the quark mass in QCD sector on the D4-branes
is shifted as
\begin{eqnarray}
 m_q \sim \frac{U_{KK}}{2\pi l_s^2}\left(r_\infty + \frac{c'}{U_0}
 \right)
 =\frac{U_{KK}r_\infty}{2\pi l_s^2}
 +\frac{g_{YM}^2\langle \bar{Q}Q\rangle}{81\pi^2 M_{KK}M_W}
\label{aaaho}
\end{eqnarray}
where we have used the W-boson mass $M_W=U_0/(2\pi l_s^2)$ of the
D4-D4$'$ string. The mass of the pion in the QCD sector is read again
\begin{eqnarray}
 m_\pi^2 \sim \frac{m_q M_{KK}}{g_{YM}^2N_c} .
\end{eqnarray}

In the above computation, we used the holographic dual description
and used the fact that
the pseudo Nambu-Goldstone wavefunction in QCD sector is
localized at $\lambda=0$ region.
We can also study the quark mass from a field theoretical point of
view, since introducing the D4$'$-branes has an interpretation of
introducing a techni gauge theory. The invariance under the rotation
in the $(x^8,x^9)$-plane is broken by the D4$'$-branes and the
breaking effects are communicated into the QCD sector on the
D4-branes by, for example, massive gauge bosons which come from the
open strings connecting the D4- and D4$'$-branes. We then have a
Feynman graph similar to Fig.\ref{w1} and can compute the quark mass
after replacing $\bar{Q}Q$ with its condensate $\langle
\bar{Q}Q\rangle$. After a careful calculation, the quark mass is
computed as a consistent value with \eqref{aaaho} up to a numerical
factor.

A comment is in order. The D6-brane configuration is obtained by
solving the equation of motion computed from the DBI action for the
D6-brane in the D4-brane geometry. This means that the pion mass
terms are induced in the DBI action because of the D4$'$-branes, on
the contrary to the Sakai-Sugimoto model where the DBI action does
not have a mass term for the pion. This is understood from the
string world sheets. The string world sheet which corresponds to the
graph in Fig.\ref{w1} is not a worldsheet instanton, but just a disk
amplitude whose boundary is on the D6-branes. The minimal worldsheet
area is vanishing. If the boundary passes through the two throats
created by the D4 and D4$'$-branes, this disk worldsheet picks up
the effects on the D4$'$-branes and communicate them into the
D4-branes. Since the DBI action is computed from  this infinitely
small disk worldsheet  (which is not a worldsheet instanton), the
mass terms for the pion appear in the DBI action in the present
case.


\section{Conclusions and Discussions}
\label{cd}

In this paper, we propose the deformations of the Sakai-Sugimoto
model to generate the quark masses. Our considerations are motivated
by extended technicolor theories where we break chiral symmetry via
introducing a technicolor sector. We systematically trace the mass
deformations in different descriptions, that is, weak coupling
D-branes setting and the corresponding holographic gravity
description. In the field theory side, the chiral condensate in the
technicolor sector is mediated to the QCD sector, generating the
quark masses. One then expects the massive quarks will yield the massive
pion. Moreover, the pion and the quark masses are expected to obey
the GOR relation from the chiral Lagrangian consideration. To {\it
derive} the GOR relation from first principles, we should have a
good control on the strong coupling dynamics of QCD. Relying on
holographic principle, our analysis verifies the GOR relation
impressively. Furthermore, we find a good numerical agreement of the
chiral condensate with the experimental or lattice results.

 In our construction, by introducing
additional technicolor branes, a novel mechanism of worldsheet
instanton gives us a controlled contribution to the quark mass
deformation. In the strong coupling regime, the $N_c$ color branes
are replaced by a curved geometry, and the corresponding worldsheet
instanton is now dressed by the Wilson line of the probe D8-brane
gauge field via worldsheet boundary interactions. This results in an
additional term in the D8-brane effective
 action corresponding to the mass deformation of the chiral Lagrangian of the pions.

We realize our idea with different types of D-branes, the D4$'$- and
$D6$-branes. The former has a clear field theory interpretation in
terms of a GUT-like extended technicolor theory, while the latter
case is more tractable in actual calculations. Moreover, it
allows us flavor-dependent quark mass terms. QCD $\theta$ angle can
also be introduced easily by turning on $C^{(2)}_{RR}$ in the
background. Based on these brane settings, we have verified that the
deformations indeed correspond to the lowest mass perturbation in
the chiral Lagrangian of pions via a novel mechanism of worldsheet
instantons. We also have the GOR relation satisfied, and from this
relation we have extracted the quark masses and the chiral
condensate of the Sakai-Sugimoto model for the first time, which
happens to be surprisingly close to the lattice QCD estimate.

 We expect that the progress made
in this paper can push the Sakai-Sugimoto model towards a more
realistic nonperturbative description of QCD. We hope it may inspire
more comparisons with the experimental or lattice QCD results. For
example, we may expect the worldsheet instanton will also affect
other hadron spectrum and their dynamics simply because
both  mesons and baryons are  excitations of the quarks. As shown in
\cite{Hong:2007kx}, there is a difficulty in fitting both the baryon
and meson spectrum well at the same time in Sakai-Sugimoto model. It
is then interesting to see if our mass deformation would help to
resolve the issue.

There are many issues
in lattice QCD or nuclear physics which call for a reliable and
calculable QCD model with non-zero quark masses. For example, the
mass shift of the pions in a finite temperature or in quark-gluon
plasma phase is of much interest. These can be done only with the
models with non-zero quark mass as the one we proposed.
The study of renormalization of the value of chiral condensate is also
interesting.
Also,
lattice computations have been done in quench approximation where
quarks are heavy. It is often stated that the Sakai-Sugimoto model
corresponds to a quenched approximation. However, it is  different
from the lattice quenching of heavy quarks, because in the original
Sakai-Sugimoto model quarks are massless. It is also hoped that our
mass deformation will help the holographic QCD to address some
issues like nuclear potential in nuclear physics, where the pion
mass will play an important role in their dynamics.

{\it Note added in proof: } While we are preparing our manuscript, there appeared
a paper \cite{Aharony:2008an} which has some overlap with ours in discussing the mass deformation by the worldsheet instantons.

\section*{Acknowledgments}

T.H. would like to thank Kazuyuki Furuuchi, Seiji Terashima and Dan Tomino. H.U.Y. would like to thank
Francesca Borzumati and High Energy Group in National Central University,
Taiwan for an invitation. K.H.~is grateful to members of String theory
group in Taiwan for hospitality, and also thanks H.~Suzuki and A.~Miwa
for valuable discussions. This work is partially supported by the Japan
Ministry of Education, Culture, Sports, Science and Technology,
Taiwan's National Center for Theoretical Sciences and National Science Council
(No. NSC 97-2119-M-002-001, NSC 96-2112-M-003-014).

\appendix

\section{Field-theoretical computation of the four-fermi term}
\label{app0}

The theory living on the D4-branes and the D4$'$-branes is 5 dimensional.
The four-Fermi coupling appears effectively after one integrates out
the massive W-bosons. The coupling between the W-boson and quarks
(techni-quarks) is given by gauge couplings
\begin{eqnarray}
\frac{g}{2} \int d\tau d^4x
\left[
\delta(\tau) \bar{q}_L \gamma_\mu W^\mu Q_L
+
\delta(\tau-\pi/M_{\rm KK}) \bar{q}_R \gamma_\mu W^\mu Q_R
\right].
\label{gaugeW}
\end{eqnarray}
Note the delta-functions are inserted in the coupling. The $N_f$
D8-branes are located at $\tau=0$ while the $N_f$ $\overline{\rm
D8}$-branes are located at the anti-podal point, $\tau = \pi/M_{\rm
KK}$. On the D8-D4 intersection, the quark $q_L$ lives, while on the
D8-D4$'$ intersection, the quark $Q_L$ lives. As for the
intersection with the $\overline{\rm D8}$-branes, $q_R$ and $Q_R$
live in the same manner. This zero-mode condition for the quarks is
represented as the delta-functions above.

The coupling (\ref{gaugeW}) is
gauge invariant and also invariant
under the chiral flavor symmetry $U(N_f)_L\times U(N_f)_R$.
The coefficient $g$ is
the gauge coupling of the
5-dimensional Yang-Mills theory on the D4-branes and the D4$'$-branes,
and so it is related to the 4-dimensional Yang-Mills
coupling constant by a dimensional reduction,
$(2\pi/M_{\rm KK})({1}/{g^2}) =  {1}/{g_{\rm YM}^2}$,
that is to say, the KK zero mode along the direction $\tau$ is
the gluon in 4-dimensions.

To derive the effective four-Fermi coupling, we
decompose the
5-dimensional notation (\ref{gaugeW}) into 4-dimensional fields,
following the standard prescription given
by \cite{Mirabelli:1997aj}. We just employ a KK expansion of all
the fields and compute the expanded theory as if it were a theory of
infinite number of 4-dimensional fields.
The KK expansion for the W-boson is given with the periodic boundary
condition as\footnote{We neglect the component $\mu=\tau$ of the W-boson
for simplicity. It is just equivalent to an adjoint scalar field, with
which the computations presented below will follow in the same manner.}
\begin{eqnarray}
 W_\mu(x,\tau) = \sum_{n=0}^{\infty}\left[
\sin(n\tau M_{\rm KK}) W_\mu^{(n)}(x)
+ \cos(n\tau M_{\rm KK}) \tilde{W}_\mu^{(n)}(x)
\right].
\label{expw}
\end{eqnarray}
{}From this expression, it is obvious that
the mass of the
$n$-th KK W-boson field $W_\mu^{(n)}$ and
$\tilde{W}_\mu^{(n)}$ is given by
$ (m^{(n)})^2 = M_W^2 + n^2 M_{\rm KK}^2$.
When we substitute the expansion (\ref{expw})
to the coupling (\ref{gaugeW}), we obtain two couplings
\begin{eqnarray}
\frac{g}{2}
\sum_n \int d^4x \; \bar{q}_L \gamma^\mu \tilde{W}_\mu^{(n)} Q_L,
\quad
\frac{g}{2} \sum_n (-1)^n\int d^4x \; \bar{Q}_R \gamma^\mu
\tilde{W}_\mu^{(n)}
 q_R.
\label{co2}
\end{eqnarray}
In the latter coupling, the factor $(-1)^n$ comes from $\cos(\pi n)$
which indicates that this coupling resides on the point
$\tau = \pi/M_{\rm KK}$.

Next, from these couplings, we integrate out the massive W-bosons.
By Wick-contracting the $\tilde{W}$ field in the couplings (\ref{co2}),
we obtain a four-Fermi coupling,
\begin{eqnarray}
-2\int d^4x
\int d^4y
\int \frac{d^4p}{(2\pi)^4}\left[
\frac{M_{\rm KK}}{2\pi}
\frac{1}{p^2 + M_W^2}
+
 \frac{M_{\rm KK}}{\pi}
\sum_{n=1}^\infty \frac{(-1)^n}{p^2 + M_W^2 +n^2 M_{\rm KK}^2}
\right]
\nonumber \\
\times e^{ip\cdot(x-y)}\times g^2
\bar{q}_L(x) q_R(y) \bar{Q}_R(x) Q_L(y).\qquad\qquad
\label{coupling2}
\end{eqnarray}
We have used a Fierz identity.
Note that the first factor $M_{\rm KK}/(2\pi)$ comes as a
normalization of the kinetic term of the zero mode in the KK expansion,
$\int d\tau\;  1^2 = {2\pi}/{M_{\rm KK}}$,
while the factor $M_{\rm KK}$ in the second term comes similarly as
$\int_0^{2\pi/M_{\rm KK}}\!\! d\tau\; \cos^2(n\tau M_{\rm KK})
= {\pi}/{M_{\rm KK}}.$
In (\ref{coupling2}), let us assume a large $M_W$,
so the momentum can be neglected, $-\partial^2 \equiv
p^2\ll M_W^2$.
Then we get the expression
\begin{eqnarray}
-2 \left[
\frac{M_{\rm KK}}{2\pi}
\frac{1}{M_W^2}
+
 \frac{M_{\rm KK}}{\pi}
\sum_{n=1}^\infty
\frac{(-1)^n}{M_W^2+ n^2 M_{\rm KK}^2}
\right]
\int d^4x \; g^2\bar{q}_L(x) q_R(x) \bar{Q}_R(x) Q_L(x).
\end{eqnarray}
Using the formula
\begin{eqnarray}
 \sum_{n=1}^\infty \frac{(-1)^n}{n^2 + s^2} = -\frac{1}{2s^2}
+ \frac{\pi}{2s} \frac{1}{\sinh s\pi},
\end{eqnarray}
The four-Fermi coefficient is evaluated as
\begin{eqnarray}
-g^2 \frac{1}{M_W} \frac{1}{\sinh(\pi M_W/M_{\rm KK})}
\int d^4x \; \bar{q}_L(x) q_R(x) \bar{Q}_R(x) Q_L(x).
\end{eqnarray}

When the techni-quarks $Q$
on the D4$'$-branes get condensed to form $\langle \bar{Q}_L Q_R\rangle$,
this four-Fermi term generates a mass term for the quarks.
In our approximation, the effectiveness requires
$M_W\gg M_{\rm KK}$. In this parameter region,
the quark mass is obtained as\footnote{
In the evaluation we drop trivial overall numerical factors.}
\begin{eqnarray}
 m_q =
\frac{g^2\langle \bar{Q}_R Q_L\rangle}{2M_W} \exp\left[
-\frac{\pi M_W}{M_{\rm KK}}
\right]
=\frac{\pi g_{\rm YM}^2
\langle \bar{Q}_R Q_L\rangle}{M_{\rm KK}M_W} \exp\left[
-\frac{\pi M_W}{M_{\rm KK}}
\right].
\label{qmass}
\end{eqnarray}
The exponential factor is important, it coincides with our worldsheet
instanton calculation.

\section{Curved shape of the probe D6-brane}
\label{appA}

In the flat spacetime background, the probe D6-brane is flat.
But once we replace the $N$ D4-branes by its geometry,
the worldvolume of the D6-brane is curved due to the
force between the D6-branes and the D4-branes,
so that it is stable.
Let us calculate this stable configuration.
We introduce spherical coordinates for the directions transverse to the
D4-brane,
\begin{eqnarray}
 x^5 = U \cos\theta_1,
\quad x^6 = U \sin\theta_1 \cos\theta_2, \cdots
\end{eqnarray}
To avoid difficulties in computation,
we make an approximation that the D6-brane
configuration is almost the same as that of the D6-brane not ending on
the D8-brane but wrapping the $x^4$ circle.
We choose the following ansatz for the D6-brane
worldvolume parameterized by
$(\sigma_1, \sigma_2, x^{0,1,2,3,4})$,
\begin{eqnarray}
 \theta_3=\theta_4=0, \quad \theta_2=\sigma_2, \quad
\theta_1 = \sigma_1, \quad U=U(\sigma_1)
\end{eqnarray}
which is maximally rotation-symmetric.
The D6-brane effective action, which is a 7-dimensional
DBI action in the Witten's background spacetime,
is given by
\begin{eqnarray}
 S = {\cal T}_{\rm D6} \int d\sigma_1 d\sigma_2 d^5x
e^{-\phi} \sqrt{-\det g}
 =
V \int\! d\theta_1\; U^{5/2}\sin\theta_1 \sqrt{
\left(\frac{dU}{d\theta_1}\right)^2 + U^2 f(U)
},
\end{eqnarray}
where $V = R^{-3/2}{\cal T_{\rm D6}}\int d^5x$.
We can easily solve the equation of motion of this action numerically,
with the initial condition
\begin{eqnarray}
 U(\theta_1=0) = U_0, \quad
 \frac{dU}{d\theta_1}\biggm|_{\theta_1=0} = 0.
\end{eqnarray}
Here $U_0$ is a constant parameter. This $U_0$
is the minimum distance between the
D6-brane and the D4-branes.
Our numerical result shows a consistent configuration of a curved
D6-brane in the background. The
worldvolume point $\theta_1=0$ is the closest to the D4-branes.
It is the same as the flat configuration of the D6-brane in the flat
background.

\section{Two-instantons and $\pi^0$-$\pi^{\pm}$ mass difference}
\label{appB}

The two instanton sector is described in the following way.
The instanton number is just the wrapping number of the boundary of the
worldsheet on the non-trivial cycle on the D8-brane.
When it winds once, the instanton contribution is proportional to
(\ref{phia}). So, the two instanton sector is given by
\begin{eqnarray}
 \sum_i\sum_j
  {\rm tr}\left[
{\rm P}\exp \left[-\Phi \int_{b} dz\right]
\cdot {\rm P} \exp\left[-i\oint_i A_z dz \right]
\cdot {\rm P}\exp \left[-\Phi \int_{b} dz\right]
\cdot {\rm P} \exp\left[-i\oint_j A_z dz \right]
\right]
\end{eqnarray}
Here, again we have followed the approximation that the integral regions
of $\Phi$ is separated from the integral region of $A_z$.
Then,
using (\ref{qmf}) and looking at the front factor in the one-instanton
result (\ref{ws1}), we obtain
\begin{eqnarray}
 S_{\rm 2-instanton} = \frac{1}{3^{9/2}\pi^3}
g_{\rm YM}^3 N_c^{3/2}M_{\rm KK}^3{\cal N}^{-2}
\cdot \frac{g_{\rm YM}^2}{2\pi M_{\rm KK}}
\int d^4x\;
{\rm tr} \left[
m_q U m_q U +  U^\dagger m_q U^\dagger m_q
\right]
\label{ws2}
\end{eqnarray}
This expression is, again, consistent with chiral perturbation
theory. We expand this expression with
$U = \exp [2i \pi(x)/f_\pi]$, then we obtain terms quadratic in
the pion fields as
\begin{eqnarray}
&&
{\rm tr} \left[
m_q U m_q U +  U^\dagger m_q U^\dagger m_q
\right]_{{\cal O}(\pi^2)}
 =
-\frac{8}{f_\pi^2} {\rm tr}
\left[
m_q \pi(x)m_q \pi(x) + m_q^2 \pi(x)^2
\right]
\nonumber
\\
&&
=
-\frac{8}{f_\pi^2}
\left((m_u^2 + m_d^2)(\pi^0(x))^2
+ \frac12(m_u+m_d)^2 ((\pi^+(x))^2 + (\pi^-(x))^2))
\right)
\nonumber
\end{eqnarray}
where $\pi^{\pm}\equiv (\pi_1 \pm i\pi_2)/\sqrt{2}$ and $\pi^0 = \pi_3$.
{}From this, the mass difference is obtained as
\begin{eqnarray}
|m^2_{\pi^\pm}-m^2_{\pi^0}|
= \frac{2}{3\sqrt{3}}\frac{g_{\rm YM}^3}{\sqrt{N_c}{\cal N}^2} 
(m_u-m_d)^2.
\end{eqnarray}

Let us give a numerical estimate as an illustration.
If we substitute $N_c=3$, $g_{\rm YM}=2.35$ and also the experiment
values of the pion masses, 
$m_{\pi^\pm} = 139.6 \;{\rm [MeV]}$, $m_{\pi^0} = 135.0 \;{\rm [MeV]}$, 
then with ${\cal N}=1$, we obtain $|m_u-m_d|\simeq 21 \; {\rm [MeV]}$.
This is larger than the experimental values of the quark masses.

The origin of this discrepancy can be understood as follows. In a chiral
perturbation theory, there are other higher order terms,
$({\rm tr} [m_q U])^2$ and $({\rm tr} [m_q U^\dagger])^2$.
Together with these
terms, the realistic pion mass difference is reproduced.
In our holographic QCD approach, these double-trace operators
appear at a sub-leading order in the large $N$ expansion and so does not
appear in our leading-order estimates.
It would be interesting to compute these sub-leading corrections and
see how the masses of $K^{0,\pm}$ can be reproduced from strange quark
mass.

\end{document}